\documentclass{mem}
\usepackage{natbib}\usepackage{txfonts}\usepackage{balance}
\usepackage{flushend}
\usepackage{graphicx}
\usepackage[breaklinks,pdftex]{hyperref}
\idline{75}{282}
\begin{document}
\def\teff{$T\rm_{eff }$}
\def\kms{$\mathrm {km s}^{-1}$}
\newcommand{\icarus}{Icarus}
\newcommand{\psj}{Planet. Sci. J.}

\title{Windows Into Other Worlds}

\subtitle{Pitfalls in the physical interpretation of exoplanet atmospheric spectroscopy}

\titlerunning{Windows Into Other Worlds}

\author{
D. \,Modirrousta-Galian\inst{1} 
\and R. \, Spinelli\inst{2} \and A. Jimenez-Escobar\inst{2}}

\institute{
Yale University, Department of Earth and Planetary Sciences, 210 Whitney Avenue, New Haven, CT 06511, USA; \email{darius.modirrousta-galian@yale.edu}\\
\and
INAF - Osservatorio Astronomico di Palermo, P.za Parlamento 1, 90134 Palermo, Italy\\
\email{riccardo.spinelli@inaf.it, antonio.jimenez@inaf.it}\\}

\date{~~~~~~~~~~~~~~~~~~~~~~~~~~~~~~~~~~~~~~~~~~~~~~~~~~~~~~~~~~~~(Accepted)}

\abstract{Atmospheric spectroscopy provides a window into the properties of exoplanets. However, the physical interpretation of retrieved data and its implications for the internal properties of exoplanets remains nebulous. This letter addresses three misconceptions held by some atmospheric spectroscopists regarding the connection between observed chemical abundances and theory: (1) Whether atmospheric spectroscopy can provide the bulk atmospheric chemistry, (2) whether it can identify if a planet is cloudless, and (3) whether atmospheric evaporation arguments can be used to dismiss certain compositions inferred through spectroscopy. This letter concludes by exploring applications of remote sensing in the quest for the search for life outside of our solar system.

}
\maketitle{}

\section{Introduction}

As our knowledge of exoplanets continues to grow, researchers are increasingly exploring detailed questions about the physical traits and formation processes of these distant worlds. Yet, analyses suffer frequently from strong degeneracies because of data limitation and our incomplete understanding of the connection between observed data and underlying physical mechanisms. The geophysical and geodynamical properties of exoplanets remain among the least understood aspects of this field, mainly because data is usually limited to mass and radius measurements, occasionally supplemented by albedo \citep{Essack2020,Modirrousta2021} and tidal Love numbers \citep{Batygin2009,Kramm2012}. One of the most promising approaches to constrain the properties of exoplanets is atmospheric spectroscopy \citep{Burrows2014,Madhusudhan2019}, which can be incorporated into geophysical analyses to reduce degeneracies \citep{Dorn2017}. Data from space telescopes like Hubble, Spitzer, and the James Webb Space Telescope, along with upcoming missions such as Ariel and Twinkle, provide opportunities for data scientists to analyze large exoplanet spectroscopic datasets. Interpreting the physical meaning behind this data presents its own challenges, but such challenges are not always appreciated in the recent literature. This letter aims to clarify common misconceptions by explaining relevant concepts clearly to help avoid them.

This letter examines three commonly held assumptions in atmospheric spectroscopy: Whether atmospheric spectroscopy can provide the bulk atmospheric chemistry (Section~\ref{sec:bulk_composition}), whether it can identify if a planet is cloudless (Section~\ref{sec:clouds}), and whether atmospheric evaporation arguments can dismiss certain compositions inferred through spectroscopy (Section~\ref{sec:rejection}); it concludes by discussing how atmospheric characterization may be used in the search for extraterrestrial life.

\section{Can we determine the bulk atmospheric composition?}
\label{sec:bulk_composition}

We obtain atmospheric compositional information from retrieval and emission spectra. Some researchers then use this data to reach conclusions about the global compositional structure of the atmosphere, such as it being hydrogen-rich (i.e., primordial). However, such claims sometimes lack nuance because they overlook many details that obscure our ability to make such general comments. For example, the upper atmosphere of Earth \citep{Emmert2021}, as well as that of Uranus and Neptune \citep{Garcia2018}, are hydrogen-rich, yet neither host hydrogen-rich atmospheres. The reason for this disparity is because planetary atmospheres are often chemically stratified. A major source of chemical stratification is cloud formation, which is discussed further in section~\ref{sec:clouds}. Above its respective cloud layer, a condensible species cannot exist because it would condense and rain down. Thus, clouds act as a chemical barrier that inhibit our ability to see the full composition of an atmosphere.

Another source of chemical stratification is the efficacy of mixing, which is not equal at all altitudes. To understand how mixing occurs in an atmosphere, we introduce the concepts of molecular and eddy diffusion. Molecular diffusion describes the gradual dispersion of individual molecules by collisions resulting from their thermal motions. It operates in all directions independently of the magnitude of the bulk flow. Eddy diffusion, also known as turbulent diffusion, arises from eddies that mechanically mix the fluid. Unlike molecular diffusion, it can exhibit directionality and it can depend on the magnitude of the bulk flow. Because molecular diffusion depends on the thermal velocity of particles, which depends on their molecular mass, different species experience different molecular diffusion rates. Thus, molecular diffusion has the ability to chemically fractionate a fluid. In contrast, eddy diffusion, being a macroscopic process, is less dependent on the mass of the species, and it can therefore result in chemical homogenization. The magnitude of molecular and eddy diffusion is proportional to their respective diffusivity coefficients, denoted as $D$ and $K_{\rm zz}$ respectively. It follows that in sections of the atmosphere where $K_{\rm zz}{>}D$, gas is well mixed (called the homosphere), whereas in sections where $K_{\rm zz}{<}D$ (called the heterosphere), differential separation occurs and chemical gradients exist.

\begin{figure}[htbp]
    \centering
    \includegraphics[width=1\linewidth]{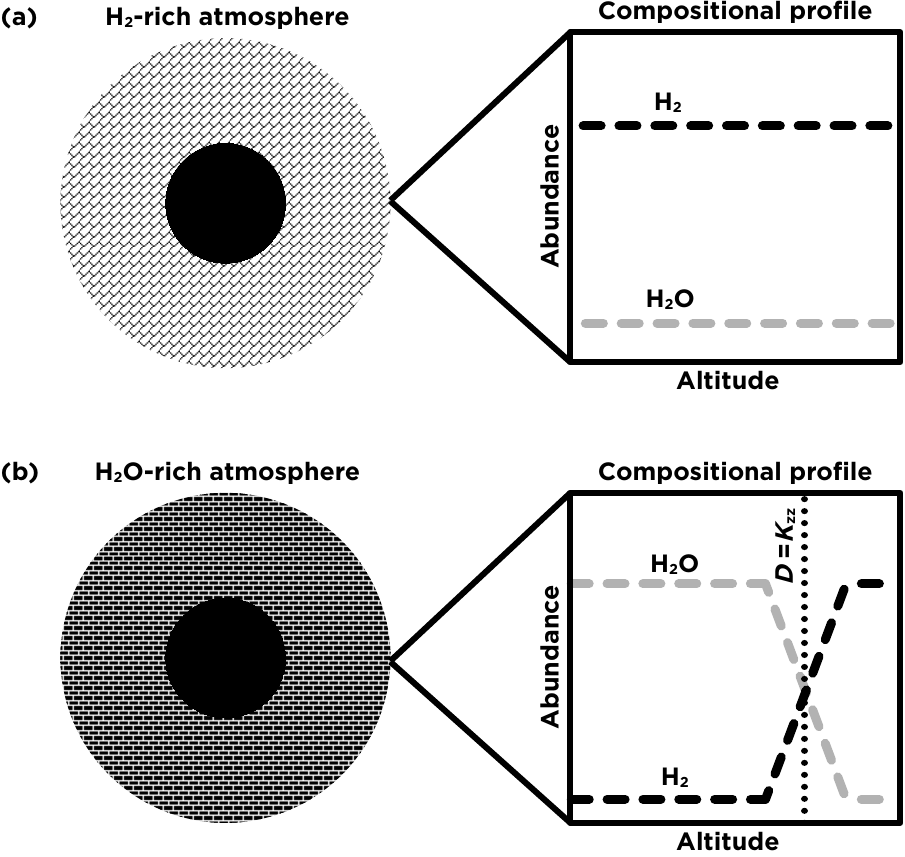}
    \caption{Schematic diagram showing the compositional profiles of two idealized rocky planets, with the first hosting a H$_{2}$-rich atmosphere and the second hosting a H$_{2}$O-rich atmosphere. The planet with the H$_{2}$-rich atmosphere is assumed to have $K_{\rm zz}{>}D$ throughout whereas the planet with the H$_{2}$O-rich atmosphere has a homosphere and a heterosphere. Both planets have H$_{2}$-rich upper atmospheres.}
    \label{fig:cartoon}
\end{figure}

We gain clearer insights into the effects of diffusion by examining an idealized rocky planet with a H$_{2}$O-rich atmosphere and another with a H$_{2}$-rich atmosphere (Figure~\ref{fig:cartoon}). Hydrogen has the highest molecular diffusion coefficient of all the elements in the periodic table, and it is thus able to diffuse through the heterosphere more rapidly than other species. This leads to hydrogen-rich upper atmospheres, as observed for Earth, Uranus, and Neptune. Thus, whether a planet is H$_{2}$-rich matters little so long as hydrogen is sufficiently abundant to populate the upper sections of its atmosphere (Figure~\ref{fig:cartoon}).

Chemical stratification may also occur from double-diffusive convection, which has been suggested to apply to the interior of Jupiter \citep{Leconte2012,Nettelmann2015,Moll2017} and Uranus and Neptune \citep{Markham2021}. This describes convection in the presence of a compositional gradient that acts against the effects of thermal buoyancy; it is less efficient than standard thermal convection in mixing and it can result in compositional stratification \citep{Huppert1981,Stern1998,Radko1999}. A detailed review of double-diffusive convection falls outside the scope of this letter and can be found elsewhere in the literature \citep[e.g.,][]{Rosenblum2011,Mirouh2012,Wood2013}. In other words, there are strong degeneracies resulting from unequal mixing that can sometimes hinder our ability to discern the bulk atmospheric composition of exoplanet atmospheres.

\section{Can we determine if a planet is cloudless?}
\label{sec:clouds}

\begin{figure}[htbp]
    \centering
    \includegraphics[width=1\linewidth]{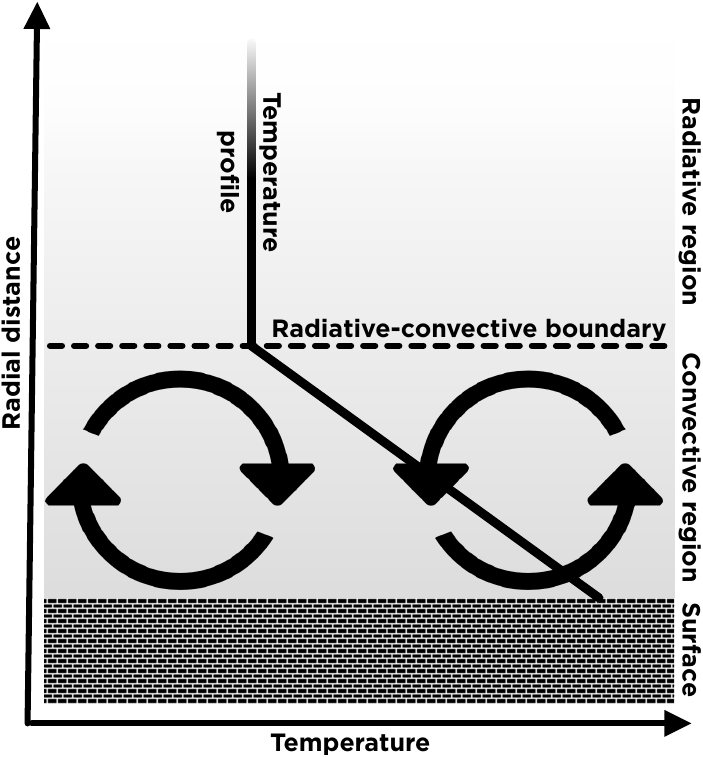}
    \caption{Schematic diagram showing the standard temperature profile of an arbitrary atmosphere. Sections below the radiative-convective boundary are convective whereas those above are radiative.}
    \label{fig:cartoon_2}
\end{figure}

\begin{figure*}[htbp]
    \centering
    \includegraphics[width=1\textwidth]{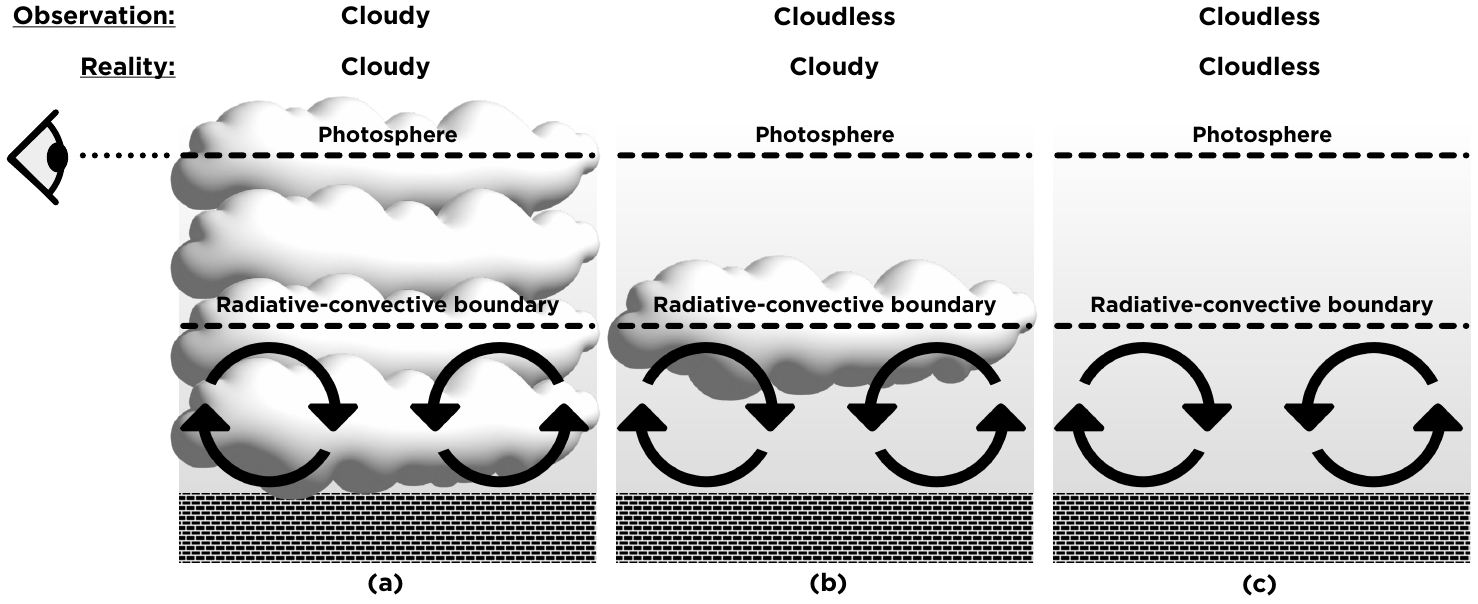}
    \caption{Schematic diagram showing three possible scenarios for a cloudy atmosphere. Case (a) corresponds to a fully cloudy atmosphere in which observations and reality agree.  Case (b) corresponds to an atmosphere with clouds at the radiative-convective boundary only, leading observations to suggest a cloudless atmosphere, which is incorrect. Case (c) corresponds to a completely cloudless atmosphere where observations and reality agree. The eye symbolizes the observable region of the atmosphere.}
    \label{fig:cartoon_3}
\end{figure*}

Clouds mask spectroscopic signatures and may therefore lead to flat observed spectra. Thus, when spectroscopic data is flat, the planet is assumed to host clouds, whereas when it is not, it is assumed to be cloudless. In reality, however, we cannot tell if a planet is cloudless, and this is explained in what follows. Clouds are composed of small droplets of a condensible fluid suspended in a non-condensed medium (e.g., air) and, thus, to understand cloud physics one has to understand condensation. An arbitrary species x condenses when its partial pressure,
\begin{equation}
    P_{\rm x} = f_{\rm x} P,
\label{eq:Px}
\end{equation}
is greater than its vapor pressure,
\begin{equation}
    P_{\rm v,x} = P_{0}\exp{\left(-\frac{L_{\rm x}}{RT}\right)},
\label{eq:Pv}
\end{equation}
where $f_{\rm x}$ is the mole fraction of species x, $P$ is the total pressure, $P_{0}$ is a constant, $L_{\rm x}$ is the latent heat of vaporization, $R$ is the ideal gas constant, and $T$ is the temperature. From Equations~\ref{eq:Px} and \ref{eq:Pv}, we see that cloud formation is favored when ambient pressure $P$ is high (i.e., large $P_{\rm x}$) and temperature $T$ is low (i.e., small $P_{\rm v,x}$). We can now consider these arguments in the context of a standard characteristic atmosphere (Figure~\ref{fig:cartoon_2}). 
The lowest section of an atmosphere is convective, so its temperature approximates an adiabat \citep{Modirrousta2023}. Because temperatures increase from the radiative-convective boundary downward, vapor pressure increases exponentially and cloud formation is mitigated in this direction. Above the radiative-convective boundary, the atmosphere is approximately isothermal \citep{Hubeny2003,Guillot2010} while pressure decreases approximately exponentially \citep{Lente2020}, sharply lowering the partial pressure and also mitigating cloud formation. Consequently, cloud formation is disfavored below and above the radiative-convective boundary, making the boundary itself the most favorable location for clouds to form. Of course, thermodynamic properties differ among species, introducing variability in the location where they condense. The key insight, however, is that condensible species are most likely to form clouds near the radiative-convective boundary, which lies deeper than the locations observable through spectroscopic analyses. Astronomical instruments can only probe regions close to the photosphere (optical depth $\tau{\approx}2/3$), whereas the radiative-convective boundary may have $\tau{\sim}10^{4}$ or more for a highly irradiated planet. In other words, the lack of a flat spectrum does not imply a cloudless atmosphere because clouds may exist at much deeper altitudes (Figure~\ref{fig:cartoon_3}).

Claims of cloud signatures observed in the spectra of hot Jupiters may initially seem contradictory to the arguments presented above. However, this apparent contradiction can be addressed by examining the energy balance of irradiated exoplanet atmospheres. In fluids, heating from below promotes convection whereas heating from above leads to stable stratification by creating a buoyant hot upper layer. Consequently, it could be argued that in highly irradiated exoplanets, the radiative-convective boundary should exist at a lower altitude because intense irradiation inhibits convection at higher altitudes. It therefore appears paradoxical why clouds are observed when the radiative-convective boundary is suggested to be at deeper regions.
\begin{figure}[htbp]
    \centering
    \includegraphics[width=1\linewidth]{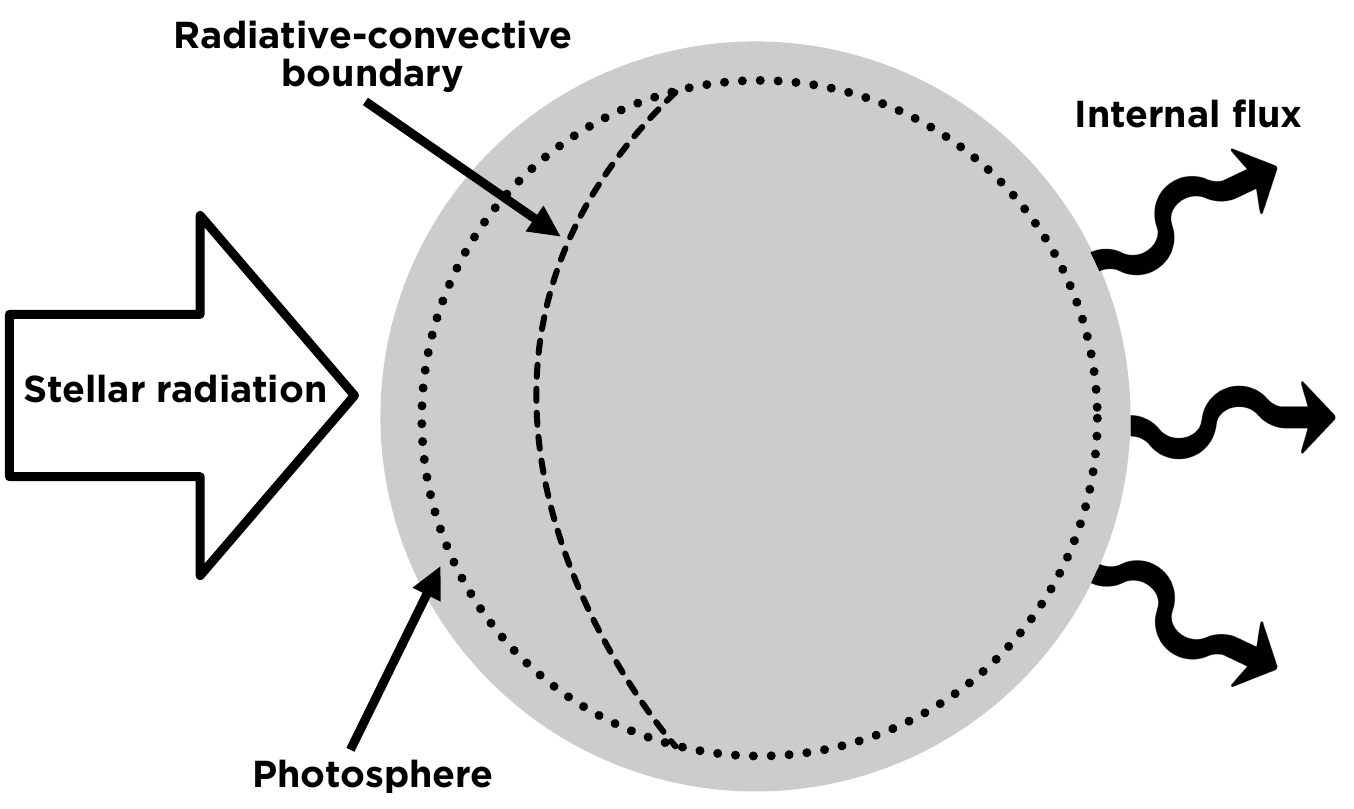}
    \caption{Cartoon showing the photosphere ($\tau{\sim}2/3$) and radiative-convective boundary of a highly irradiated planet that has a non-negligible internal heat flux. In the day-side, the radiative-convective boundary is at a much lower depth than the photosphere, whereas at the terminator and the night-side, they converge.}
    \label{fig:cartoon_4}
\end{figure}

Addressing this conundrum first requires the recognition that observations probe regions close to the terminator of the planet, where incoming stellar flux is much reduced (Figure~\ref{fig:cartoon_4}). The radiative flux arriving at a given latitude is approximately $F(\phi){=}F_{0}\cos(\phi)$, where $F_{0}$ is the substellar point flux and $\phi$ is the latitude probed. The position of the radiative-convective boundary is governed by the equilibrium between the incoming stellar radiation and the interior flux so that, for example, if we probe latitudes $\phi{>}70^{\circ}$, the flux arriving is $F(70^{\circ})/F_{0}{\sim}0.35$. The first part of this apparent paradox is therefore resolved, that is, highly irradiated hot Jupiters are not as highly irradiated at their terminators. The next part is to recognize that hot Jupiters have high internal heat fluxes, which is known from observations suggesting that they have radii larger than that expected from modeling \citep[see review;][]{Dawson2018}. Whereas the mechanism or mechanisms responsible for their radial inflation still elude us, statistical analyses indicate nonetheless the necessity of very high internal luminosities \citep{Thorngren2018,Thorngren2019,Sarkis2021}. 

The reduced stellar flux at the terminator with the high internal luminosities required to inflate hot Jupiters brings the radiative-convective boundary close to the photosphere, enabling cloud visibility through observations. This configuration is unlikely to apply to smaller exoplanets because they have lower internal heat fluxes, highlighting the importance of being nuanced when interpreting exoplanet spectroscopic data.

\section{Can we dismiss some compositions based on atmospheric evaporation?}
\label{sec:rejection}

The last assumption discussed in this letter is the suggested dismissal of certain atmospheric compositions inferred through spectroscopy, based on whether they are compatible with atmospheric evaporation arguments. This reasoning applies mostly to primordial hydrogen-rich atmospheres, which are more prone to being lost because hydrogen has the lowest mass and the fastest thermal velocity of all the elements. Consider, for example, a spectroscopic analysis suggesting the detection of a hydrogen-rich atmosphere on a hot super-Earth or sub-Neptune. Indeed, one may be suspicious because hot small-mass planets are unlikely to host hydrogen stably because of their low gravities and high temperatures. However, there are several exoplanets that appear to defy this notion by seemingly hosting primordial atmospheres while closely orbiting their stars, such as NGTS-4~b \citep{West2019}, LTT~9779~b \citep{Jenkins2020},
and TOI-908 b \citep{Hawthorn2023}. Moreover, new analyses suggest that hydrogen loss from X-ray and ultraviolet irradiation is inefficient because momentum diffusion between different species slows the escape rate of hydrogen \citep{Modirrousta2024}. We also do not know the initial hydrogen reservoir of the planet, which could have been sufficiently large \citep{Rafikov2006,Rafikov2011,Ikoma2012,Lee2015} to survive stellar irradiation. Last, some authors have suggested that hydrogen could be stored within planetary interiors and released later over geological timescales \citep{Chachan2018,Schlichting2022}, protecting it from the extreme environments when the star \citep{Penz2008(1),Penz2008(2)} and planet \citep{Abe1997} are young. In short, we cannot yet conclusively rule out certain atmospheric compositions based solely on atmospheric evaporation arguments because our understanding of the physics is incomplete. Further research is needed to fully understand the complexities of exoplanetary atmospheres and their evolution.

\section{Discussions and application to the search for life}

Understanding the physical interpretation of exoplanet spectroscopic data is necessary to evaluate their atmospheric and surface environments, which is crucial for assessing their potential habitability. Evidence of complex chemistry can be detected on a distant planet through remote sensing of its atmosphere, and we discuss this possibility in the sections that follow.

\subsection{Electron discharge chemistry in exoplanetary atmospheres}
\label{sec:discharge}

Electrical discharges occur regularly in Earth's atmosphere, and they can modify the composition of air by increasing its chemical complexity. These processes can generate the building blocks of life such as amino acids (section~\ref{sec:lab}), which play crucial roles in protein synthesis, enzyme functions, and the development of living organisms. Such electrical processes are common in the solar system, and they have been observed in the atmospheres of Jupiter \citep{Cook79}, Saturn \citep{Warwick81}, and Uranus and Neptune \citep{Zarca86,Gurnett90,Kaiser91}. Indeed, electrical discharges are also expected on Mars, where they have been suggested to form from frictional electrification of dust in its atmosphere \citep{Krauss03}. Therefore, if we assume that exoplanets bare similar characteristics to the solar system planets, we would expect them to also experience electrical discharges and potentially form some of the building blocks to life.

\subsubsection{Laboratory experiments}
\label{sec:lab}

\begin{figure}[htbp]
    \centering
    \includegraphics[width=1\linewidth]{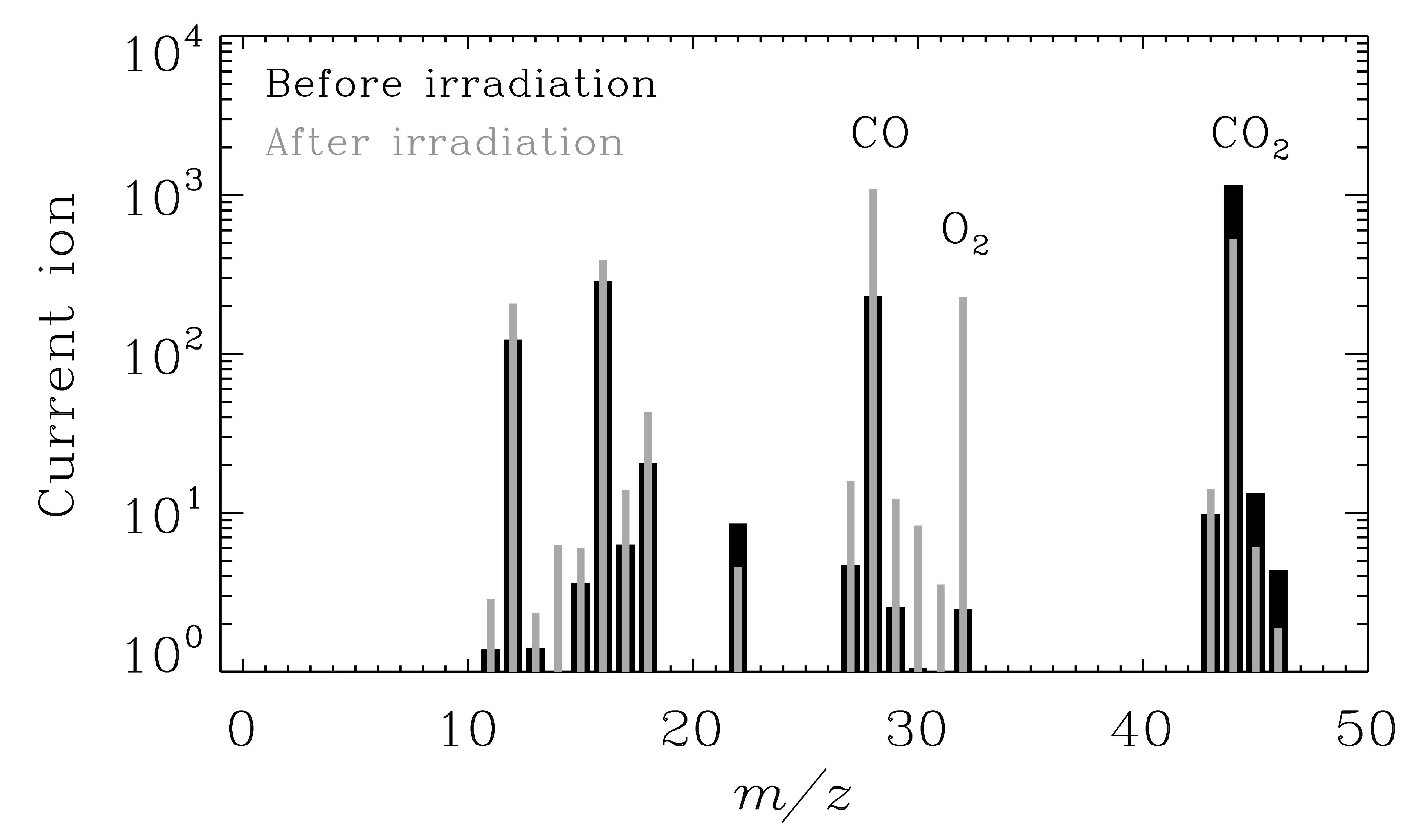}
    \caption{Mass spectrum of pure CO$_2$ gas before (black line) and after (grey line) irradiation \citep{jimenez23}.}
    \label{fig:co2}
\end{figure}

Experiments of electrical discharge processing were conducted to assess non-equilibrium chemistry in CO$_2$-rich atmospheres, resembling that of present-day Mars and the Hadean Earth \citep{Kasting93,Delano01}. Gas was exposed to a high voltage, and its evolving composition was monitored using mass spectrometry, revealing a rapid decrease in CO$_2$ and the detection of O$_2$ and CO as primary gas phase products (Figure~\ref{fig:co2}). 
These experiments suggest that molecular oxygen, crucial for life, can form through abiotic processes. These experiments were also found to produce a dusty solid residue rich in -OH, C=C, and C-O bearing organic species. Such organic dust may be generated during lightning, later being incorporated into hazes and clouds that are detectable through spectroscopy. 

\subsection{Ultraviolet Habitable Zone}

Recent experiments suggest that prebiotic photochemistry leading to the formation of the building blocks of RNA requires a minimum Near Ultraviolet (NUV, 200$-$280~nm) flux threshold \citep{Patel2015,Xu2018,Rimmer2018}. These findings raise doubts on the validity of the classical Habitable Zone (HZ) description for planets around non solar type stars, such as red dwarfs who have a greater UV-to-bolometric luminosity ratio. Building on the framework of \citet{Rimmer2018}, \citet{Spinelli2023} introduced a new definition of stellar UV-Habitable Zone (UHZ) in which the inner and outer orbital distance boundaries are defined by the maximum and minimum UV flux tolerable for life and required for the emergence of life through cyano-sulfidic chemistry respectively. Their findings suggest that eighteen out of twenty-three HZ planets considered orbit outside the UHZ. It follows that for these planets the present-day NUV luminosity of their cold host star ($T_{\ast}{<}$3900~K) is insufficient to trigger abiogenesis through standard cyano-sulfidic chemical pathways. 
Further studies are necessary to assess if such chemistry was possible during the first ${\sim}100~{\rm Myr}$ after a star has formed when its NUV luminosity was at its maximum \citep{Penz2008(1),Penz2008(2)}. Moreover, we note that the origin of life is actually very poorly understood, particularly the last stages of prebiotic chemistry that lead to the emergence
of life. Notwithstanding, whether such NUV radiation can reach the planetary surface is highly dependent on the composition of the atmosphere and the luminosity of the star \citep{Cnossen2007}. By determining the atmospheric composition through spectroscopy, we may be able to evaluate its transmissivity to NUV radiation and thus infer the likelihood of such chemistry taking place. To achieve this goal, it is necessary to obtain more data and to have a thorough understanding of what information we can attain from it to avoid pitfalls that give a false sense of certainty.

\section{Conclusions}

Exoplanet science is at the cusp of a data revolution, and with this data it is necessary to be nuanced and precise with one's interpretations. Common misconceptions may slow progress by giving a false sense of certainty. Now, more than ever, data scientists and theoreticians need to collaborate together to avoid these pitfalls.
 
\begin{acknowledgements}
This work was sponsored by the US National Science Foundation EAR-2224727 and the US National Aeronautics and Space Administration under Cooperative Agreement {No.\,80NSSC19M0069} issued through the Science Mission Directorate. This work was also supported in part by the facilities and staff of the Yale University Faculty of Arts and Sciences High Performance Computing Center.
\end{acknowledgements}

\bibliographystyle{agsm}
\bibliography{bibliography}

\end{document}